\documentclass[12pt,preprint]{aastex}

\usepackage{graphicx}

\shorttitle{Transiting planet yield of Gaia}

\begin{document}

\title{Detection of transiting Jovian exoplanets by \textit{Gaia} photometry -- expected yield} 
\author{Yifat Dzigan and Shay Zucker}
\affil{Dept. of Geophysical, Atmospheric and Planetary Sciences,
	Raymond and Beverly Sackler Faculty of Exact Sciences, Tel
	Aviv University, Tel Aviv 69978, Israel}
\email{yifatdzigan@gmail.com,shayz@post.tau.ac.il}
\date{Accepted for publication in the Astrophysical Journal Letters. Accepted 2012 May 16.}

\begin{abstract}

Several attempts have been made in the past to assess the expected
number of exoplanetary transits that the \textit{Gaia} space mission
will detect.  In this Letter we use the updated design of
\textit{Gaia} and its expected performance, and apply recent empirical
statistical procedures to provide a new assessment. Depending on the
extent of the follow-up effort that will be devoted, we expect
\textit{Gaia} to detect a few hundreds to a few thousands transiting
exoplanets.

\end{abstract}


\section{Introduction} 

\textit{Gaia} is a planned European Space Agency (ESA) mission,
scheduled to be launched at 2013. It will perform an all-sky
astrometric and spectrophotometric survey of point-like objects
between 6th and 20th magnitude.  The primary goal of the telescope is
to explore the formation, dynamical, chemical and star-formation
evolution of the Milky Way galaxy. The main science product of 
\textit{Gaia} will be high precision astrometry, backed with photometry 
and spectroscopy.  It will observe about 1 billion stars, a few
million galaxies, half a million quasars, and a few hundred thousands
asteroids \citep{2010IAUS..261..296L}.

\textit{Gaia} will operate in a Lissajous-type orbit, around the L2
point of the Sun-Earth system, about $1.5$ million kilometers from
Earth in the anti-Sun direction. It will have a dual telescope, with a
common structure and common focal plane.  During its five-year
operational lifetime, the spacecraft will continuously spin around its
axis, with a constant speed of $60\,\mathrm{arcsec\,s}^{-1}$. As a
result, during a period of six hours, the two astrometric fields of
view will scan all objects located along the great circle
perpendicular to the spin axis. As a result of the basic angle of
$106.5\degr$ separating the astrometric fields of view on the sky,
objects will transit the two fields of view with a delay of $106.5$
minutes.  Due to the spin motion of six-hour period, and a
$63$-day-period precession, the scanning law will be peculiar and
irregular.  This scanning law will result in a total of $70$
measurements on average for each celestial object \textit{Gaia} will
observe \citep{2012Ap&SS.tmp...68D}.

\textit{Gaia} will provide photometry in several passbands, the widest
of which will be a 'white' passband dubbed $G$, centered on $\lambda_0
= 673\,\mathrm{nm}$, with a width of $\Delta\lambda =
440\,\mathrm{nm}$.  In what follows we use the apparent $G$ magnitude
as approximately equal to the apparent $V$ magnitude. One can expect a
milli-magnitude (mmag) precision in the $G$ band for most of the 
objects \textit{Gaia} will observe, down to 14th--16th $G$ magnitude, and 
$10\,\mathrm{mmag}$ at the worst case of 19th magnitude objects 
\citep{2010A&A...523A..48J}. The exact limiting magnitude for a $1\,\mathrm{mmag}$ 
precision depends on instrumental factors which are not yet ``frozen'' \citep{2012Ap&SS.tmp...68D}. 

The $1\,\mathrm{mmag}$ precision of \textit{Gaia} photometry naturally
raises the question of whether it can be used to detect exoplanetary
transits.  While $1\,\mathrm{mmag}$ precision is nominally more than
sufficient for the detection of Jovian transiting planets, the low
cadence and the small number of measurements make the feasibility of
this detection a non-trivial question. In the literature, there have
been several conflicting estimates as to the number of transits
detectable by \textit{Gaia}, based on different assumptions.

\cite{2002Ap&SS.280..139H} estimated the expected number of transit
detections by \textit{Gaia} at $250$, for long-period planets, with
orbital radius of $0.5\,\mathrm{AU}$. According to
\cite{2002Ap&SS.280..139H}, the expected yield of hot Jupiters (HJs)
and very hot Jupiters (VHJs), with orbital radii of
$0.1-0.01\,\mathrm{AU}$ was $6250$ and $625000$ respectively.
\cite{2002Ap&SS.280..139H} based these estimates on some general
assumptions. First, the planets frequency was approximated as
$\sim1\%$, which already proved as overestimated for HJs
\citep[][hereafter BG]{2008ApJ...686.1302B}.  Second,
\citeauthor{2002Ap&SS.280..139H} did not account for the stellar
density and its variation due to galactic structure, and neglected
extinction by dust. He assumed that the number of stars \textit{Gaia}
will observe with $1\,\mathrm{mmag}$ precision (up to magnitude
$G=15.5$) is $10^9$. Finally, he assumed that a transit detection can
be made with only one transit observation per system, counting on
\textit{Gaia} astrometric observations to complement the transit
observation.
 
\cite{2002EAS.....2..215R} performed transit simulations with the
assumed \textit{Gaia} photometry to estimate the number of detections,
and concluded that \textit{Gaia} will detect between $4000$ and
$40000$ transiting Jupiter-like planets. He assumed that the number of
individual observations per star would be between $100$ and $300$ with
an average of $130$. He also used a galactic model
\citep{1997A&A...320..440H}, and derived the probability distribution
of the number of observations during transits using \textit{Gaia}
scanning law.  We suspect that these predictions are overestimated,
due to the fact that the currently planned scanning law of
\textit{Gaia} implies an average of $70$ measurements per star, and
not $130$.  We will refer to \citeauthor{2002EAS.....2..215R}'s
estimates in more detail in section \ref{results}.

In this short Letter we present a new estimate, one that we believe is
more realistic than the previous ones. It is based on the methodology
of BG, which includes broad assumptions about the galactic structure,
as well as implicit assumptions about the geometric transit
probability and the effects of stellar variability, that rely on
statistics from completed transit surveys. The field of planetary
transits of HJs and VHJs seems to have come to a certain maturity from
which we can draw some statistical assumptions. We feel it is not yet
the case for transits of smaller planets ('Saturnian' and
'Neptunian'), and we therefore do not address these issues here.

Providing more accurate and up-to-date predictions of the number of
transiting planets detectable by \textit{Gaia} is important for the
ongoing effort of developing the \textit{Gaia} analysis
pipeline. Moreover, detection of planetary transits requires
considerable follow-up effort, which is also a reason for having a
reliable estimate of this number.  We took upon ourselves to provide a
somewhat more rigorous analysis, based on empirical statistics, simply
because the estimates in the literature are too varied and
inconsistent.  At the time they were made, the field of transit
surveys was too young, and much had yet to be learned. The time
has come to provide a more decisive estimate, based on the current
more evolved understanding of the problem.

\section{Predicting the transit yield}

BG presented a statistical methodology to predict the yield of
transiting planets from photometric surveys.  Their method takes into
consideration the frequency of short period planets, variations in the
stellar density due to the galactic structure and also corrects for
the extinction by dust.

Following the procedure suggested by BG, the average number of
exoplanets that a photometric survey can detect is estimated by the
product of the probability to detect a transiting planet, the
frequency of transiting planets and the local stellar mass
function. Obviously, this product should also be integrated over mass,
distance and the field of view:

\begin{equation}\label{formalism1}
 \frac{d^6N_\mathrm{det}}{dR_\mathrm{p}\,dp\,dM\,dr\,dl\,db} = \rho_{\ast}(r,l,b)\,r^2 \cos{b} \frac{dn}{dM} \frac{d^2f(R_\mathrm{p},p)}{dR_\mathrm{p}\,dp}  P_\mathrm{det}(M,r, R_\mathrm{p},p),
\end{equation}
$\rho_{\ast}(r,l,b)$ is the local stellar density as a function of
heliocentric galactic coordinates $(r,l,b)$, $dn/dM$ is the present
day mass function in the solar neighborhood, and
$\frac{d^2f(R_\mathrm{p},p)}{dR_\mathrm{p}\,dp}$ is the frequency of
transiting planets (the probability that a given star harbors a
transiting planet with radius $R_\mathrm{p}$ and period $p$).
$P_\mathrm{det}$ is the transit detection probability, assuming 
there is indeed a transiting planet around the examined star.

Following BG we consider detection of transits of Very Hot Jupiters
(VHJs) with orbital periods of $1-3$ days, and Hot Jupiters (HJs) with
periods of $3-5$ days.  For the probability that a star will harbor a
transiting planet with radius $R_p$ and orbital period $p$ we assume
the same form that BG used:
\begin{equation}
 \frac{d^2f(R_\mathrm{p},p)}{dR_\mathrm{p}\,dp} = k(p)\,f(p)\,\delta(R_p-R^{'}_p).
\end{equation}

The normalization factor $k(p)$ is an empirical number which can be
deduced from completed surveys. It encapsulates many factors which are
common to transit surveys looking for the same range of periods and
more or less the same stellar populations.  BG proposed to use a
normalization factor based on the results of
\cite{2006AcA....56....1G}, who analyzed the statistics of the OGLE
transit surveys. They suggested a value of $k(p)=1/690$ for VHJs, and
$k(p)=1/310$ for HJs, and a locally uniform distribution of the
period in the specified interval, $f(p)=1$.

This planet frequency \citep{2006AcA....56....1G} 
is compatible with the frequency which is implied by Kepler results \citep{2011arXiv1103.2541H}.
Furthermore, \textit{Gaia}'s low cadence makes it more similar to ground-based surveys, 
rather than to high-cadence space surveys. Moreover, 
CoRoT and Kepler results are not yet complete, since they are still operating.

To account for the stellar density in the solar neighborhood, we
used the Present Day Mass Function (PDMF).
\cite{2002AJ....124.2721R} used data from Palomar/Michigan State
University survey, together with the Hipparcos dataset, to derive the
PDMF:
\begin{equation}
\frac{dn}{dM} = \left\{
	\begin{array}{ll}
	k_\mathrm{norm}(\frac{M}{M_\odot})^{-1.35} & \textrm{for $0.1\leq M/M_\odot \leq 1$ } \\
	k_\mathrm{norm}(\frac{M}{M_\odot})^{-5.2} & \textrm{for $1< M/M_\odot$ }
	\end{array} \right.
\end{equation}
where again we adopted the normalization suggested by BG,
$k_\mathrm{norm}=0.02124\,\mathrm{pc}^{-3}$.

In order to convert absolute magnitudes to masses, we used the
mass-luminosity relations, as derived by
\cite{2002AJ....124.2721R}. For lower main-sequence stars we used the
mass-luminosity relation:
\begin{equation}
 \log M = 10^{-3}\times(0.3+1.87M_V+7.614M^2_V-1.698M^3_V+0.06096M^4_V) ,
\end{equation}
and for the upper main-sequence stars, 
\begin{equation}
\log M = 0.477-0.135M_V+1.228\times10^{-2}M^2_V-6.734\times10^{-4}M^3_V ,
\end{equation} 
where $M_V$ is the absolute visual magnitude of the star. The boundary
between the calibrations is set at $M_V=10$.

The last step of the procedure is integration of the stellar density
over the entire field of view:

\begin{equation}
 N_\mathrm{det}=\int_0^{\infty} \int_{l_\mathrm{min}}^{l_\mathrm{max}} \int_{b_\mathrm{min}}^{b_\mathrm{max}}   
\rho_{\ast}(r,l,b)\, r^2 \cos {b}\,db\,dl\,dr .
\end{equation}

We now had to account for the variation in the stellar density due to
the galactic structure, and also the effect of extinction due to
interstellar dust. We incorporated into our calculations the galactic
model of \cite{1980ApJS...44...73B}, which is a simplified model that
depends only on the distance from the galactic plane ($z$) and the
disk-projected distance from the galactic center ($d$):
\begin{equation}
 \rho_{\ast} = \exp \left[-\frac{d}{h_{\mathrm{d},*}}- \frac{|z|}{H(M_V)} \right] ,
\end{equation}
$h_{\mathrm{d},*}=2.5\,\mathrm{kpc}$ is the scale length of the disk.
The scale height, $H(M_V)$,
depends on the absolute magnitude, and BG use the dependence: 
\begin{equation}
 H(M_V) = \left\{ \begin{array}{lll} 90 & \textrm{for $M_V \leq 2 $ }
 \\ 90+200 \left ( \frac{M_V-2}{3} \right ) & \textrm{for $2 < M_V <5$
 } \\ 290 & \textrm{for $ M_V \geq 5 $ } \end{array} \right. .
\end{equation}

To account for the interstellar dust extinction, we used the
expression suggested by \cite{1980ApJS...44...73B} for obscuration by
dust. In their model the obscuration depends on the heliocentric
distance and the galactic latitude:
\begin{equation}
A(r) = \left\{
	\begin{array}{ll}
	0 & |b| > 50^0  \\
	0.17\,(1.2+\tan{|b|})\, [1-\exp(-\frac{r\sin{|b|}}{h}) ] \,\csc{|b|} & |b| \leq 50^0 
        \end{array} \right. ,
\end{equation}
where $h$ is a typical scale height with respect to the galactic plane,
 that is approximated by $100\,\mathrm{pc}$ \citep{1980ApJS...44...73B}. 

The term $P_\mathrm{det}(M,r, R_\mathrm{p},p)$ in Eq.\
\ref{formalism1} deserves special attention.  In their original
formalism, BG followed the common wisdom and assumed that the
detection probability is mainly a function of the transit
signal-to-noise ratio (SNR), which is usually defined by:
\begin{equation}\label{snr}
 \mathrm{SNR}=\sqrt{n_\mathrm{tr}} \frac{\Delta}{\sigma},
\end{equation}
where $n_\mathrm{tr}$ is the number of observed transits, $\Delta$ is
the transit depth, and $\sigma$ is the photometric error. In BG's
treatment, $\sigma$, and therefore the SNR, strongly depend on the
stellar magnitude, which is obviously the case in most surveys.

However, unlike in most surveys, an important feature in the design of
\textit{Gaia} is a 'gating' mechanism, that will cause most bright
stars, certainly down to the $16$th magnitude, to be measured with a
precision of $0.001$ mag, more or less
\citep{2010A&A...523A..48J}. Thus, the SNR in our case is mainly a
matter of the number of observations that occurred during transits.

BG assumed that a planetary transit can be detected whenever the
transit SNR exceeds a threshold. However, as opposed to high-cadence
surveys, due to the small number of measurements that will sample the
transits, Eq.\ \ref{snr} does not represent the complete detection
problem in our case. Even if the SNR does exceed a threshold value, we
still cannot regard the signal we found as a periodic one.  In fact,
in our assumed parameters, using the SNR na\"{i}vely, even a single
observation in transit can be considered a transit detection with a
SNR of $10$. Obviously, we have to use another indicator, which mainly
depends on the number of points observed in transit, and abandon the
SNR figure of merit for our purposes (that would not be the case for
smaller planets, though, where an even more elaborate indicator will
probably be needed). To put it differently, we assume the problem is
not limited by the error-bars of the individual measurements, but only
by the scanning law and the temporal characteristics of the transit.

\cite{2009ApJ...702..779V} calculated, for various ground surveys,
the detection probability as a function of the period, which they
dubbed the 'observational window function'. In our ideal case we
assumed pure white noise, and no outliers. We further assumed a
certain transit duration, and a minimum number of points in transit
that would constitute a detection (depending on the detection approach
used).  We then calculated for each period, the fraction of
configurations (namely, transit phases) which will result in
detection, i.e., when the number of observations in transit will
exceed the prescribed minimum.  Once we had obtained the observational
window function, we could integrate it over the required period range
and obtain an estimate of the detection probability.

The minimum number of observations in transit required for detection
is not clear at the moment. 
\cite{2011A&A..529A...6T} introduced an algorithm, that requires 
a minimum number of seven to eight points in transit to secure
detection. In an upcoming paper (Dzigan \& Zucker, in preparation),
we show that we can detect transits with five points in transit, or 
use our 'Directed Follow-Up' approach \citep{2011MNRAS.415.2513D}
even for three points in transit. We therefore repeated our 
calculations here for a minimum number of three, five and seven points in transits.

\begin{figure}[ht!]
\includegraphics[width=\textwidth]{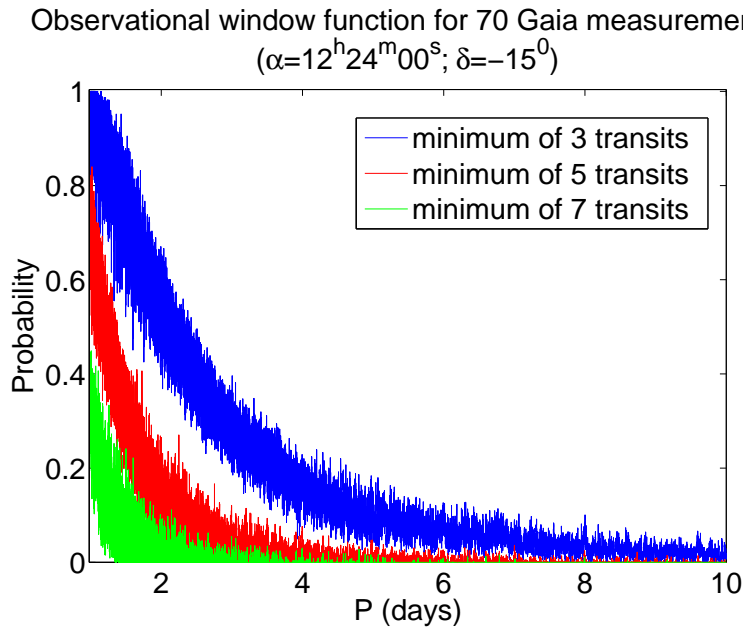}
\caption{A sample observational window function for a sky direction
that \textit{Gaia} is expected to observe $70$ times, the average
expected number of measurements over the entire mission. The detection
probability is calculated for transit duration of $2$ hours, for a
minimum of three, five and seven observations in transit.}
\label{fig.window1}
\end{figure}

We divided the entire sky into rectangular patches, $15\degr \times
15\degr$ (apart for the poles, obviously, where we simply used the
remaining circular patch), and applied the proper scanning law for
each patch, assuming the scanning law for the central point as
representing the entire patch.  Fig.\ \ref{fig.window1} shows a sample
observational window function for one of the patches, for three cases
of minimum points in transit (three, five and seven), and for a
transit duration of $2$ hours. This specific window function
represents an area that
\textit{Gaia} is expected to visit $70$ times. This is the average
expected number of measurements over the entire mission
\citep{2012Ap&SS.tmp...68D}.  

\begin{figure}[ht!]
\includegraphics[width=\textwidth]{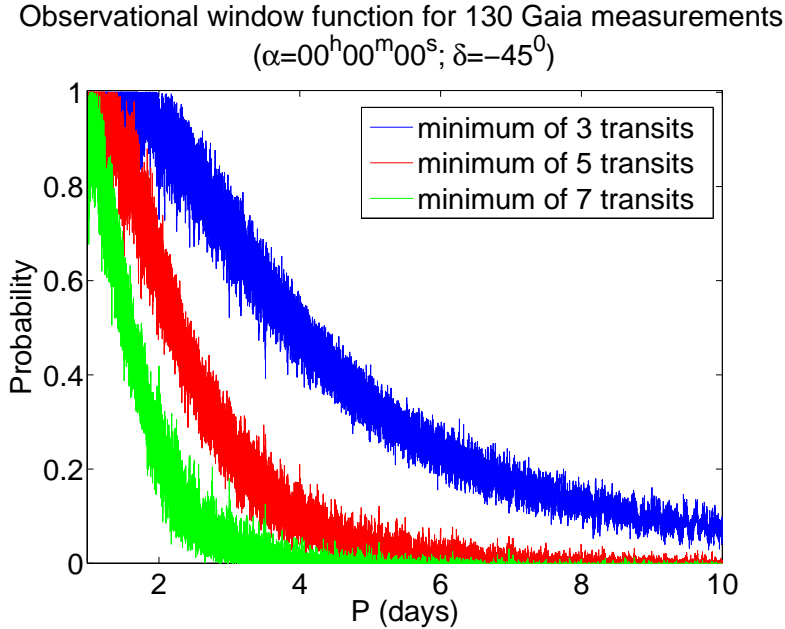}
\caption{A sample observational window function for an area that
\textit{Gaia} is expected to observe $130$ times.}
\label{fig.window2}
\end{figure}

For comparison we also present the
window function for an area with $130$ measurements in Fig.\
\ref{fig.window2}.  The comparison shows that the detection
probability depends strongly on the number of observations that the
telescope will perform, during the mission lifetime. For example, the
probability to sample a minimum of three transits (for an orbital
period of $3$ days) increases from less than $30\%$ for $70$
measurements, to more than $60\%$ in case the telescope should observe
the star $130$ times.

Since we neglected the dependence of the window function on the SNR
and therefore on the stellar characteristics, it remains mainly a
function of the period and the scanning law. We can therefore take it
out of the integral sign in Eq.\ \ref{formalism1}, which we calculate
separately for each patch. We also divided the apparent magnitude
range ($G=6-16$) into $1$-mag bins and treated each bin
separately. Only at the end of the integration we multiplied the
result with the detection probability we obtained from the window
function.

\section{Results}\label{results}

\begin{figure}[ht!]
\includegraphics[width=1\textwidth]{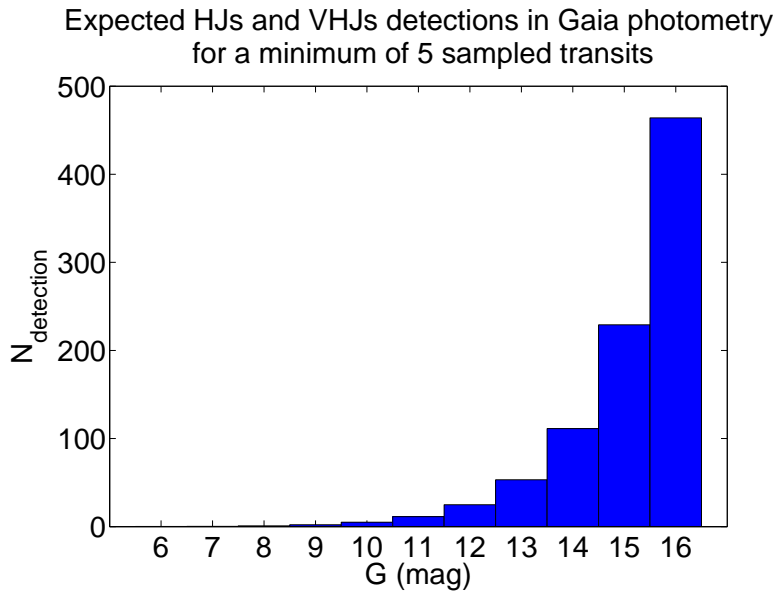}
\caption{The expected yield broken into apparent magnitude bins. 
This yield was calculated for a minimum of five sampled transits, and
a transit duration of $2$ hours.}
\label{fig.detection2}
\end{figure}

Table \ref{table.detectZ} presents the resulting expected yield of
transiting HJs and VHJs in \textit{Gaia} photometry, down to stellar
magnitudes $14$ and $16$, for which we expect \textit{Gaia} to observe on the order of $1.5\times10^5$ 
and $6.8\times10^5$ stars, respectively. 
Obviously, the required minimum number of
observations in transit strongly affects the results, as well as the
assumed transit duration.  In Fig.\ \ref{fig.detection2} we present
the transiting planets yield for a minimum of five sampled transits,
for transit duration of $2$ hours, divided into apparent magnitude
bins, from $M=6$ to $M=16$. 
Our results show that the \textit{Gaia} photometry is expected to
yield on the order of hundreds or thousands of new planets, depending
on the detection strategy.

\begin{deluxetable}{cccc}
\tablecaption{\textit{Gaia} expected detections down to a limiting
magnitude \label{table.detectZ}}
\tablehead {
\colhead{} & 
\colhead{Minimum Number of} & 
\colhead{} & 
\colhead{} 
\\
\colhead{} & 
\colhead{Points in Transit} & 
\colhead{$G=14$} & 
\colhead{$G=16$} 
}
\startdata 
\hline
                   & 3 & 230 & 999  \\
$w=1\,\mathrm{hr}$ & 5 &  42 & 178  \\
                   & 7 &   7 & 30   \\
\hline
                   & 3 & 596 & 2605 \\
$w=2\,\mathrm{hr}$ & 5 & 209 & 902  \\
                   & 7 &  73 & 310  \\
\hline   
                   & 3 & 720 & 3191 \\
$w=3\,\mathrm{hr}$ & 5 & 364 & 1577 \\
                   & 7 & 156 &  669 \\
\enddata
\tablecomments{The expected yield of HJs and VHJs from \textit{Gaia}
photometry, for three different transit durations, down to a limiting
apparent magnitudes of $G=14$, and $G=16$.}
\end{deluxetable}

Given the large amount of stars that \textit{Gaia} will measure, 
we must estimate the false-alarm rate.
The probability that $1\,\mathrm{mmag}$ purely white noise will produce 
a single 'transit-like' outlier (with magnitude of $0.005-0.015$ mag) 
is $\sim3.5\times10^{-7}$, which amounts to $\sim2\times10^{-15}$ for three outliers out of the 
average $70$ measurements. This is a worst case estimate, as requiring the outliers to have a 
periodic pattern, even reduces this probability. 
Thus, under our nominal assumptions, it is obvious that the false-alarm rate is negligible. 

Nevertheless, the analysis might be complicated by stellar red noise.
If the red noise doesn't possess a periodic or quasi-periodic nature,
then the low-cadence sampling simply renders it ``white'', effectively reducing the SNR. 
According to \cite{2012A&A...539A.137M} we estimate that roughly half of the stars have microvariability
larger than 2 mmag. This results in an increased false alarm rate. However, in our upcoming paper
(Dzigan \& Zucker, in preparation) we show that our prioritization process in the Directed Follow-Up approach 
effectiviely eliminates them.

In any case, \textit{Gaia} will provide the astrometric and spectroscopic data needed for 
further classification. These data will help to exclude false-positives, such as 
background eclipsing binaries (BGEB), and to distinguish between periodic variability of stellar source 
and planetary transits. Thus, we can conclude that we don't foresee a significant false-positive rate,
as long as we focus our analysis on HJs.

Our results seem to differ considerably from those of
\citet{2002EAS.....2..215R}. We suspect that the main cause for the
discrepancy is the different scanning law we used, which implied $50-200$ 
observations per star, with a mean of $70$, compared to
$100-300$ with a mean of $130$, which \citeauthor{2002EAS.....2..215R}
used. This reflects changes in the mission design during the years
that elapsed since 2002. Fig.\ \ref{fig.window1} and Fig.\
\ref{fig.window2} hint at the significant effect this change had on
the observational window function.

It is important to stress again that the analysis we present did not
consider smaller planets. Their detection will be more difficult, and
moreover, the required follow-up, either photometric or spectroscopic,
will be more complicated. This topic will require a much more
elaborate and careful analysis.

The analysis we present in this Letter is a rough estimate that is
based on general assumptions. We obviously made some
approximations on the way, but at the level of accuracy needed at 
this stage we feel they are justified. 
The most important conclusion is that it 
will be worth while to develop detection
algorithms, that will be tailored for \textit{Gaia} photometry, and
that will be incorporated into the pipeline. This may reduce the
 minimiun number of points in transit required for detection, which
will immensely affect the yield. 
In addition, establishing a follow-up 
network that will be able to respond to alerts will also have a 
crucial effect, again, through this reduction in the number of 
required observations in transit \citep[e.g.][]{2011arXiv1112.0187W}.

A significant feature one can notice while examining Table
\ref{table.detectZ} and Fig.\ \ref{fig.detection2} is the very strong
dependence of the yield on the limiting apparent magnitude. Extending
the analysis to fainter magnitudes will require introduction of the
SNR into the analysis. This effort will be useless unless
high-precision radial velocities of such faint targets will be
feasible. Extremely large telescopes such as the E-ELT may allow this
kind of observations. The enormous increase in the number of detected
planets with \textit{Gaia}, if fainter stars are considered, may serve as a
justification to build high-precision radial velocity spectrographs
for those telescopes.

\textit{Gaia} will undoubtedly revolutionize astronomy in many aspects.
Nevertheless, its contribution to the field of transiting exoplanets is
usually expected to be marginal. The expected transiting planet yield
is the key factor to establish whether this field will benefit
considerably from \textit{Gaia}. Usually, transit surveys focus on
dense fields, to maximize the chances to detect transits and
effectively use their high cadence. \textit{Gaia}, on the other hand,
will be an all-sky, low-cadence survey. This kind of surveys are
usually considered irrelevant for transit searches. The estimate we
present here shows that \textit{Gaia} will have a
valuable and significant contribution also in this field, mainly due
to its high photometric precision, and in spite of its low cadence.

\section{acknowledgments}

This research was supported by The Israel Science Foundation
and The Adler Foundation for Space Research (grant No. 119/07).
We are grateful to Leanne Guy and Laurent Eyer for fruitful discussions
and for providing us with the most updated scanning law of \textit{Gaia}.
We wish to thank the referee Douglas Caldwell whose valuable comments helped
to improve this paper.

%


\end{document}